\PassOptionsToPackage{nopatch=footnote}{microtype}
\documentclass[sigconf]{acmart}
\AtBeginDocument{%
  }

\setcopyright{acmlicensed}
\copyrightyear{2026}
\acmYear{2026}
\acmDOI{XXXXXXX.XXXXXXX}
\acmConference[MobiSys '26]{The 24th Annual International Conference
  on Mobile Systems, Applications, and Services (Demo)}{June 21--25, 2026}{Cambridge, UK}
\acmISBN{978-1-4503-XXXX-X/2026/06}

\begin{document}

\title{Demo: LightGuard: Transparent WiFi Security via Physical-Layer LiFi Key Bootstrapping}

\author{Shiqi Xu}
\affiliation{%
  \institution{The Chinese University of Hong Kong}
  \city{Hong Kong SAR}
  \country{China}}
\email{xs024@ie.cuhk.edu.hk}

\author{Yuyang Du}
\affiliation{%
  \institution{The Chinese University of Hong Kong}
  \city{Hong Kong SAR}
  \country{China}}
\email{yuydu@ie.cuhk.edu.hk}

\author{Mingyue Zhang}
\affiliation{%
  \institution{The Chinese University of Hong Kong}
  \city{Hong Kong SAR}
  \country{China}}
\email{1155191597@link.cuhk.edu.hk}

\author{Hongwei Cui}
\affiliation{
  \institution{The Chinese University of Hong Kong}
  \city{Hong Kong SAR}
  \country{China}}
\email{ch021@ie.cuhk.edu.hk}

\author{Soung Chang Liew}
\affiliation{%
  \institution{The Chinese University of Hong Kong}
  \city{Hong Kong SAR}
  \country{China}}
\email{soung@ie.cuhk.edu.hk}

\renewcommand{\shortauthors}{Xu, et al.}

\begin{abstract}
WiFi is inherently vulnerable to eavesdropping because RF signals may penetrate many physical boundaries, such as walls and floors. LiFi, by contrast, is an optical method confined to line-of-sight and blocked by opaque surfaces. We present LightGuard, a dual-link architecture built on this insight: cryptographic key establishment can be offloaded from WiFi to a physically confined LiFi channel to mitigate the risk of key exposure over RF. LightGuard derives session keys over a LiFi link and installs them on the WiFi interface, ensuring cryptographic material never traverses the open RF medium. A prototype with off-the-shelf WiFi NICs and our LiFi transceiver frontend validates the design. 
\end{abstract}


\maketitle

\begin{figure}[htbp]
  \centering
  \includegraphics[width=\linewidth]{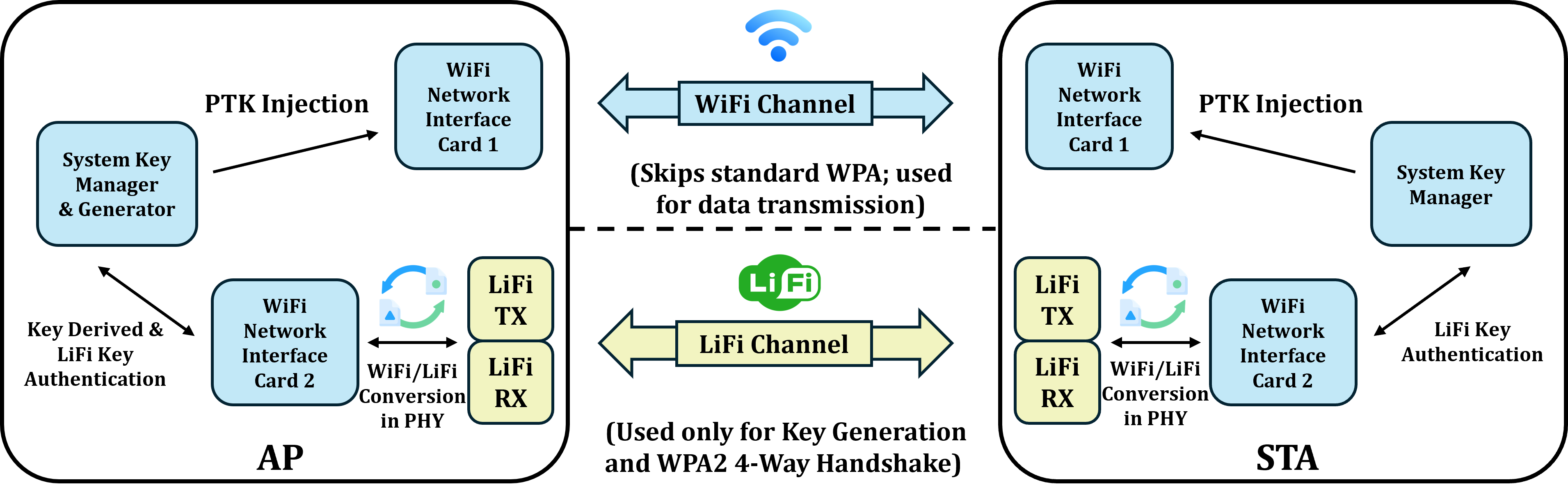}
  \caption{LightGuard dual-link system architecture: LiFi bootstraps the PTK and injects it into WiFi.}
  \label{fig:architecture}
\end{figure}

\section{Introduction}
WiFi eavesdropping remains a persistent and fundamental security challenge. A root cause is that the radio frequency (RF) medium is inherently open: any attacker within radio range can eavesdrop on all over-the-air transmissions. Practical attacks, such as passive key recovery and handshake eavesdropping, have repeatedly demonstrated the severity of this exposure \cite{liu2025survey}.

\setcounter{figure}{2}
\begin{figure*}[t]
  \centering
  \begin{minipage}[t]{0.24\textwidth}
    \centering
    \includegraphics[width=\linewidth]{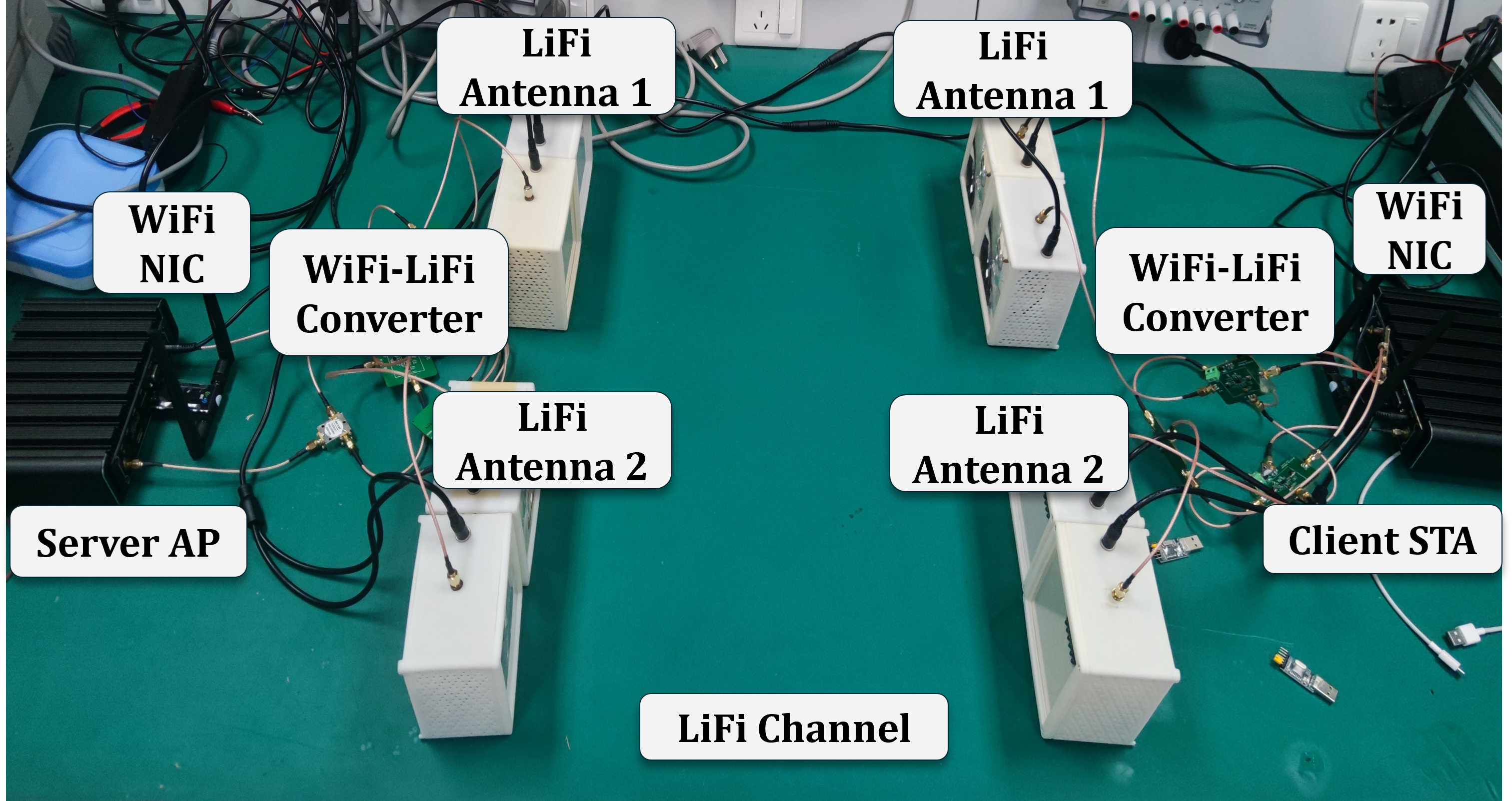}
    \caption{Real-time prototype with portable LiFi antennas.}
    \label{fig:setup}
  \end{minipage}\hfill
   \begin{minipage}[t]{0.24\textwidth}
    \centering
    \includegraphics[width=0.95\linewidth]{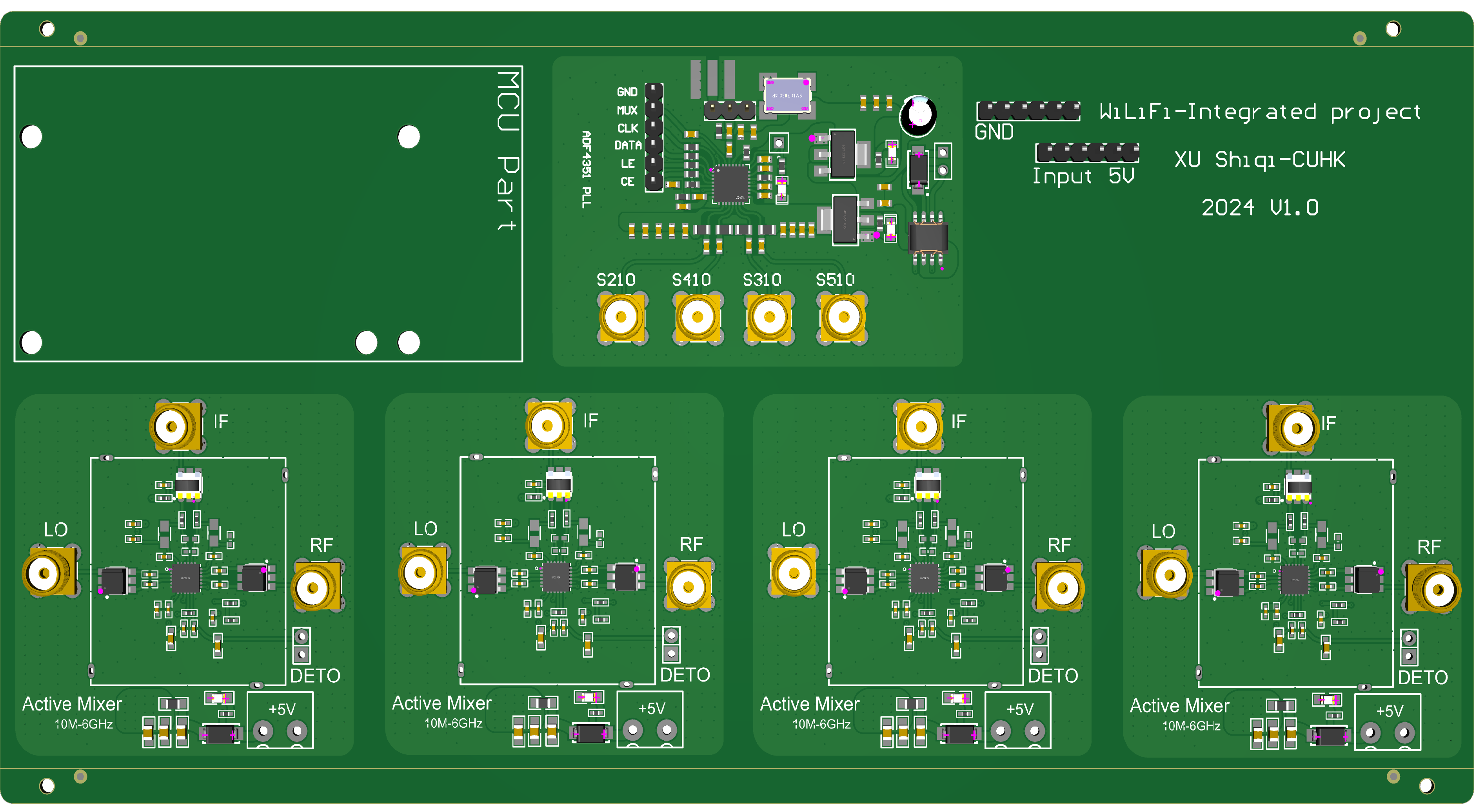}
    \caption{PCB of our LiFi antennas for WiFi/LiFi conversion.}
    \label{fig:PCB}
  \end{minipage}\hfill
  \begin{minipage}[t]{0.24\textwidth}
    \centering
    \includegraphics[width=0.93\linewidth]{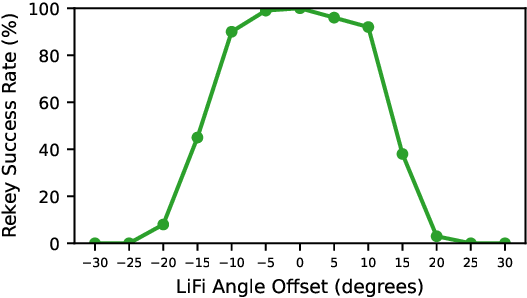}
    \caption{Rekey success rate vs LiFi angular misalignment.}
    \label{fig:throughput}
  \end{minipage}\hfill
    \begin{minipage}[t]{0.24\textwidth}
    \centering
    \includegraphics[width=\linewidth]{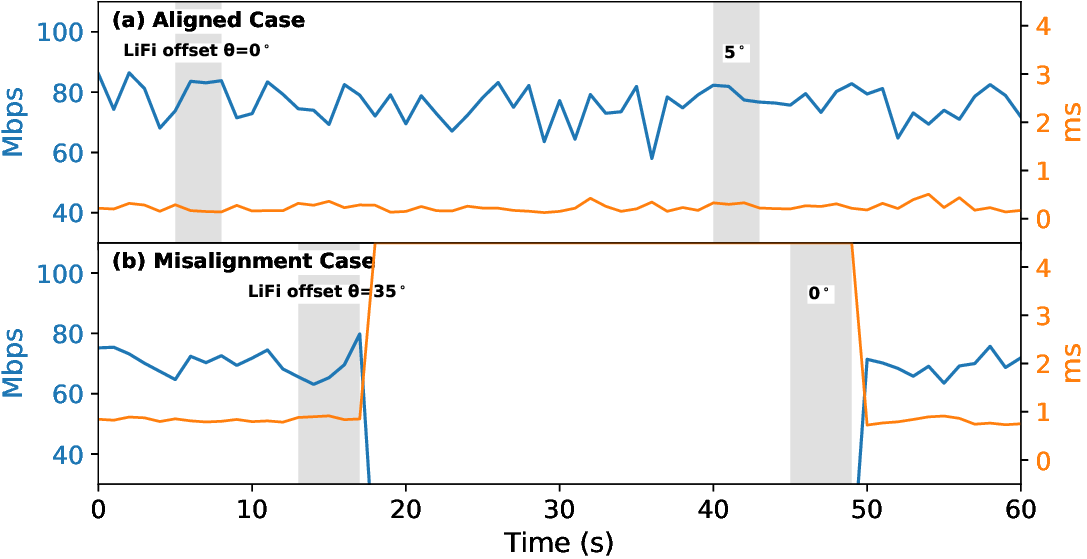}
    \caption{WiFi link under (a) aligned \& (b) misaligned LiFi.}
    \label{fig:latency}
  \end{minipage}\hfill
\end{figure*}

Light Fidelity (LiFi) offers a complementary channel whose physical properties contrast sharply with RF: visible light requires line-of-sight propagation and cannot penetrate opaque surfaces, inherently confining the transmission to a controlled physical boundary. This makes LiFi an attractive medium for security-sensitive operations: it offers stronger physical confinement than RF while retaining the deployment flexibility of a wireless link.

This paper presents \textit{LightGuard}, a hybrid dual-link architecture in which each node is equipped with both a WiFi and a LiFi network interface. The LiFi link handles cryptographic key negotiation for WiFi sessions within a physically confined optical channel, while the WiFi link carries high-throughput bulk data transfer. By offloading cryptographic key establishment entirely to the optical domain, LightGuard ensures that cryptographic material never traverses the open RF medium, making key eavesdropping infeasible for attackers outside the optical coverage area while retaining WiFi's advantages in coverage, mobility, and resilience.


We demonstrate LightGuard with commercial IEEE 802.11 NICs and novel ``LiFi antennas'' developed in \cite{cui2024wi}. These antennas perform physical-layer RF-to-optical conversion: WiFi signals are frequency-shifted onto LEDs for optical transmission and restored to RF at the receiver, allowing unmodified NICs to communicate over light without any software or protocol modification.\footnote{The Demo video is at \textcolor{blue}{\url{https://youtu.be/Qsc2zh43aZA}}}

\section{System Overview}
Figure~\ref{fig:architecture} presents LightGuard's dual-link architecture. The core idea is to offload all authentication and key-establishment traffic from the WiFi link to the LiFi link, securing the handshake without sacrificing WiFi's native throughput. Below we describe the data channel (i.e., WiFi), the security channel (i.e., LiFi), and their cross-link synchronization.

\textbf{Data Channel (WiFi):} The WiFi link carries all application traffic at native IEEE 802.11 throughput, encrypted with a Pairwise Transient Key (PTK) derived over the LiFi link. For the WiFi link, LightGuard bypasses the standard in-band over-the-air WPA2 4-Way Handshake entirely. Instead, the PTK is installed via key synchronization from LiFi.

\textbf{Security Channel (LiFi):} All cryptographic key establishment is confined to the LiFi link. The LiFi link executes standard WPA2 4-Way Handshake to derive a fresh PTK for the WiFi data channel.

\textbf{Cross-Link Key Synchronization:} Figure~\ref{fig:protocol} presents the sequence diagram of LightGuard's key synchronization process, which proceeds in four phases. In Phase 1 (Passphrase Distribution), the AP generates a fresh random passphrase and distributes it to the STA over LiFi. In Phase 2 (4-Way Handshake over LiFi), the AP and the STA execute the standard WPA2 4-Way Handshake over the optical channel to derive a shared 48-byte PTK. In Phase 3 (PTK Synchronization), the AP orchestrates a synchronized switchover via a two-phase commit. Finally, in Phase 4 (WiFi Data Resumption), both nodes inject the new PTK obtained via LiFi into their WiFi interfaces and seamlessly resume encrypted data transmission.

\setcounter{figure}{1}
\begin{figure}[htbp]
  \centering
  \includegraphics[width=0.95\linewidth]{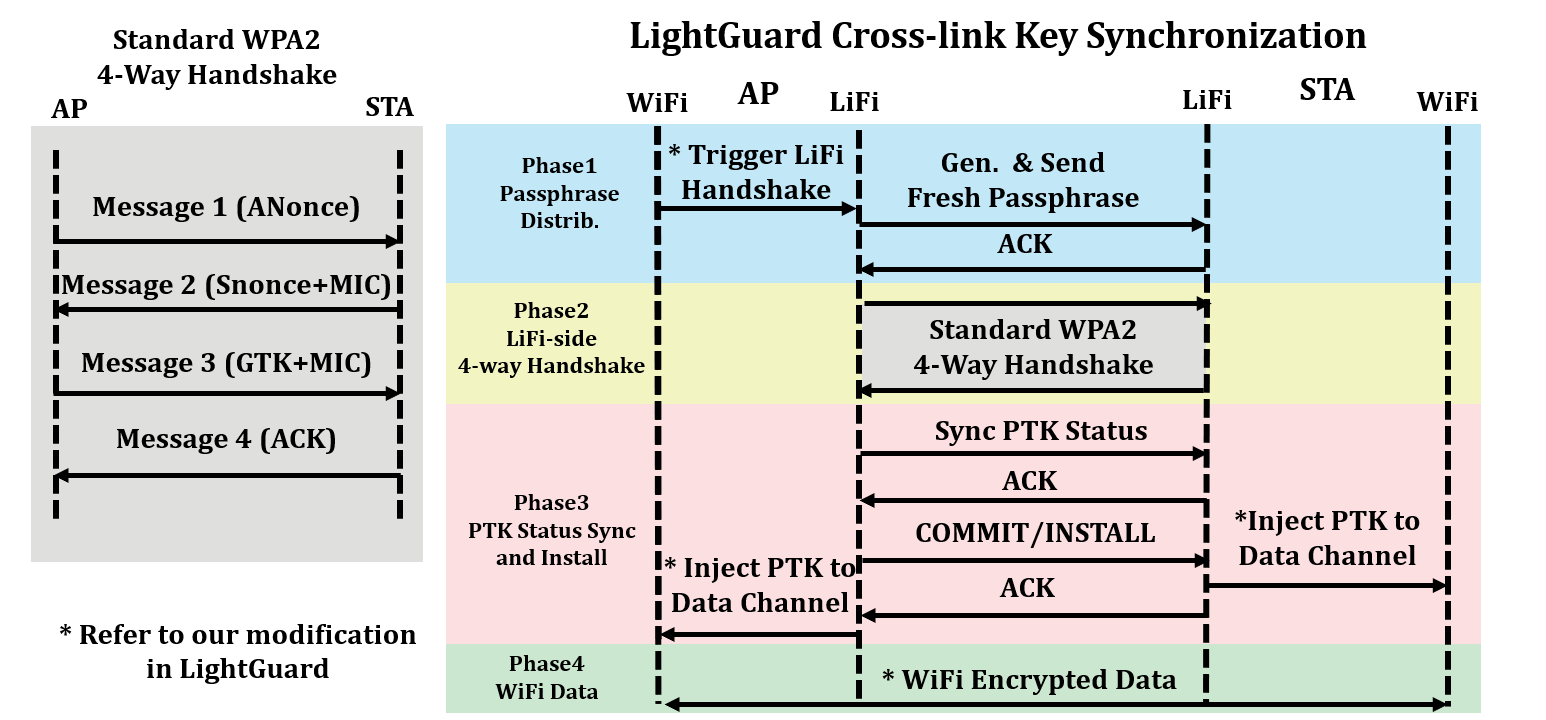}
  \vspace{-0.4cm}
  \caption{LightGuard cross-link key synchronization}
  \Description{Sequence diagram showing the four phases of LightGuard: Passphrase Distribution, 4-Way Handshake over LiFi, 2-Phase PTK Sync, and WiFi Data transmission.}
  \label{fig:protocol}
\end{figure}

\section{Implementation Details and Experiments}
Figure~\ref{fig:setup} presents LightGuard's implementation in our
lab environment. Two mini-PCs running Ubuntu 22.04 are each equipped with two Intel AX200 WiFi NICs and the LiFi transceiver developed in~\cite{cui2024wi} (see Figure \ref{fig:PCB} for the hardware design). We implemented the cross-link key synchronization process described in Figure~\ref{fig:protocol}  by extending the standard WPA2 daemons (\texttt{hostapd}/\texttt{wpa\_supplicant}) within the Linux 802.11 subsystem.\footnote{Coding details are available at \textcolor{blue}{\url{https://github.com/Dorian47/Lightguard}}.}
Both the AP and the STA run a separate \texttt{hostapd}/\texttt{wpa\_supplicant} instance for each interface. We modified the LiFi-side daemons to extract the key material  after the 4-Way Handshake; a coordinated synchronization process then forwards the PTK to both WiFi-side daemons, which install it through the standard nl80211 driver API---making each driver behave as if a local handshake had occurred. These modifications preserve full compatibility with unmodified IEEE 802.11 NICs.

As LiFi relies on directional light propagation, the transmitter and receiver must maintain line-of-sight angular alignment for reliable communication. Figure~\ref{fig:throughput} shows that rekey success is tightly coupled to LiFi alignment: the success rate peaks near boresight and degrades rapidly with increasing angular offset. Once misalignment exceeds a threshold (e.g., approximately $\pm25^\circ$ in Figure~\ref{fig:throughput}), handshake frames become unreliable and rekey attempts fail. 

Figure~\ref{fig:latency} plots WiFi throughput and latency over time, with shaded regions marking rekey phases. Under stable alignment (Figure~\ref{fig:latency}a), all rekeys succeed and the WiFi connection remains stable---throughput stays around 80\,Mbps and latency around 1\,ms. In the second experiment, when the offset is deliberately shifted to $35^\circ$ during the second rekey (Figure~\ref{fig:latency}b), the handshake fails; the system disconnects the WiFi link until realignment, after which the next rekey succeeds and the data channel recovers seamlessly. From Figure~\ref{fig:latency}, we see that even angular misalignment alone suffices to break the optical
handshake---under physical obstruction or wall separation, the isolation would only be stronger.

\section{Discussion and Conclusion}

LightGuard leverages LiFi's physical confinement to protect WiFi key establishment while preserving WiFi as the high-throughput data channel. Our prototype demonstrates that cross-link rekeying completes atomically with minimal impact on ongoing traffic, and that rekey reliability is determined by the inherent alignment constraints of the optical link.

In our current prototype, LightGuard requires the AP and STA to share line-of-sight only during brief periodic rekeying windows, while all data traffic traverses the NLoS WiFi path. To scale to larger deployments, the LiFi antennas can be extended via RF cable to serve as remote \textit{Key Stations} in access-controlled areas for cryptographic key distribution. Where Ethernet infrastructure is available, Key Stations can also leverage existing building wiring for broader deployment. Building on the LightGuard's architecture, promising next steps include evaluating multi-room coverage, investigating optimal placement strategies, and extending the rekeying process to concurrent multi-STA cryptographic key management.


\bibliographystyle{ACM-Reference-Format}
\bibliography{sample-base}

\end{document}